\newif\ifwordcount
\wordcountfalse

\documentclass[aps,prl,twocolumn,superscriptaddress,groupedaddress,titlepage,nofootinbib,amsmath]{revtex4}
\usepackage{graphicx}
\usepackage{float}

\renewcommand{\eqref}[1] {equation $($\ref{#1}$)$}
\newcommand{\comment}[1]{{}}
\def\beq{\begin{equation}}
\def\eeq{\end{equation}}
\def\beqn{\begin{eqnarray}}
\def\eeqn{\end{eqnarray}}

\def\2gcm{\textrm{g cm$^{-2}$}}

\def\H0{\ensuremath{\mathrm{H}_0}}

\def\bL{\bmm{L}}

\newcommand{\herschel}{\textit{Herschel}}
\newcommand{\Pb}{\textsc{Polarbear}}

\newcommand{\arcsec}{\ensuremath{^{''}}}

\newcommand{\arcmin}{\ensuremath{^{'}}}

\newcommand{\bmm}[1]{{\mathbf{#1}}}

\newcommand{\bm}[1]{\ensuremath{\mbox{\boldmath $#1$}}}

\def\arcsec{$^{\prime\prime}$}

\def\Bell{\bm \ell}

\begin{document}
\title{Evidence for Gravitational Lensing of the Cosmic Microwave Background Polarization from Cross-correlation with the Cosmic Infrared Background}
\author{\Pb\ Collaboration} \noaffiliation
\author{P.A.R. Ade} \affiliation{School of Physics and Astronomy, Cardiff University}
\author{Y. Akiba} \affiliation{The Graduate University for Advanced Studies}
\author{A.E. Anthony} \affiliation{Center for Astrophysics and Space Astronomy, University of Colorado, Boulder}
\author{K. Arnold} \affiliation{Department of Physics, University of California, San Diego}
\author{M. Atlas}\affiliation{Department of Physics, University of California, San Diego}
\author{D. Barron} \affiliation{Department of Physics, University of California, San Diego}
\author{D. Boettger} \affiliation{Department of Physics, University of California, San Diego}
\author{J. Borrill} \affiliation{Computational Cosmology Center, Lawrence Berkeley National Laboratory} \affiliation{Space Sciences Laboratory, University of California, Berkeley}
\author{C. Borys} \affiliation{California Institute of Technology, Pasadena, CA} 
\author{S. Chapman} \affiliation{Department of Physics and Atmospheric Science, Dalhousie University, Halifax, NS, B3H 4R2, Canada}
\author{Y. Chinone} \affiliation{High Energy Accelerator Research Organization (KEK)} \affiliation{Department of Physics, University of California, Berkeley}
\author{M. Dobbs} \affiliation{Physics Department, McGill University}
\author{T. Elleflot} \affiliation{Department of Physics, University of California, San Diego}
\author{J. Errard} \affiliation{Space Sciences Laboratory, University of California, Berkeley} \affiliation{Computational Cosmology Center, Lawrence Berkeley National Laboratory}
\author{G. Fabbian} \affiliation{AstroParticule et Cosmologie, Univ Paris Diderot, CNRS/IN2P3, CEA/Irfu, Obs de Paris, Sorbonne Paris Cit\'e, France} \affiliation{International School for Advanced Studies (SISSA)}
\author{C. Feng} \affiliation{Department of Physics, University of California, San Diego}
\author{D. Flanigan} \affiliation{Department of Physics, University of California, Berkeley} \affiliation{Columbia University}
\author{A. Gilbert} \affiliation{Physics Department, McGill University}
\author{W. Grainger} \affiliation{Rutherford Appleton Laboratory, STFC}
\author{N.W. Halverson} \affiliation{Center for Astrophysics and Space Astronomy, University of Colorado, Boulder} \affiliation{Department of Astrophysical and Planetary Sciences, University of Colorado, Boulder} \affiliation{Department of Physics, University of Colorado, Boulder}
\author{M. Hasegawa} \affiliation{High Energy Accelerator Research Organization (KEK)} \affiliation{The Graduate University for Advanced Studies}
\author{K. Hattori} \affiliation{High Energy Accelerator Research Organization (KEK)}
\author{M. Hazumi} \affiliation{High Energy Accelerator Research Organization (KEK)} \affiliation{The Graduate University for Advanced Studies} \affiliation{Kavli Institute for the Physics and Mathematics of the Universe (WPI), Todai Institutes for Advanced Study, The University of Tokyo}
\author{W.L. Holzapfel} \affiliation{Department of Physics, University of California, Berkeley}
\author{Y. Hori} \affiliation{High Energy Accelerator Research Organization (KEK)}
\author{J. Howard} \affiliation{Department of Physics, University of California, Berkeley} \affiliation{University of Oxford}
\author{P. Hyland} \affiliation{Physics Department, Austin College}
\author{Y. Inoue} \affiliation{The Graduate University for Advanced Studies}
\author{G.C. Jaehnig} \affiliation{Center for Astrophysics and Space Astronomy, University of Colorado, Boulder} \affiliation{Department of Physics, University of Colorado, Boulder}
\author{A. Jaffe} \affiliation{Department of Physics, Imperial College London}
\author{B. Keating} \affiliation{Department of Physics, University of California, San Diego}
\author{Z. Kermish} \affiliation{Physics Department, Princeton University}
\author{R. Keskitalo} \affiliation{Computational Cosmology Center, Lawrence Berkeley National Laboratory}
\author{T. Kisner} \affiliation{Computational Cosmology Center, Lawrence Berkeley National Laboratory} \affiliation{Space Sciences Laboratory, University of California, Berkeley}
\author{M. Le Jeune} \affiliation{AstroParticule et Cosmologie, Univ Paris Diderot, CNRS/IN2P3, CEA/Irfu, Obs de Paris, Sorbonne Paris Cit\'e, France}
\author{A.T. Lee} \affiliation{Department of Physics, University of California, Berkeley}  \affiliation{Physics Division, Lawrence Berkeley National Laboratory} 
\author{E.M. Leitch} \affiliation{Department of Astronomy and Astrophysics, University of
Chicago}\affiliation{Kavli Institute for Cosmological Physics, University of
Chicago}
\author{E. Linder} \affiliation{Physics Division, Lawrence Berkeley National Laboratory} \affiliation{Space Sciences Laboratory, University of California, Berkeley}
\author{M. Lungu} \affiliation{Department of Physics, University of California, Berkeley}
\author{F. Matsuda} \affiliation{Department of Physics, University of California, San Diego}
\author{T. Matsumura} \affiliation{High Energy Accelerator Research Organization (KEK)}
\author{X. Meng} \affiliation{Department of Physics, University of California, Berkeley}
\author{N.J. Miller} \affiliation{Observational Cosmology Laboratory, Code 665, NASA Goddard Space Flight Center}
\author{H. Morii} \affiliation{High Energy Accelerator Research Organization (KEK)}
\author{S. Moyerman} \affiliation{Department of Physics, University of California, San Diego}
\author{M.J. Myers} \affiliation{Department of Physics, University of California, Berkeley}
\author{M. Navaroli} \affiliation{Department of Physics, University of California, San Diego}
\author{H. Nishino} \affiliation{Kavli Institute for the Physics and Mathematics of the Universe (WPI), Todai Institutes for Advanced Study, The University of Tokyo}
\author{H. Paar} \affiliation{Department of Physics, University of California, San Diego}
\author{J. Peloton} \affiliation{AstroParticule et Cosmologie, Univ Paris Diderot, CNRS/IN2P3, CEA/Irfu, Obs de Paris, Sorbonne Paris Cit\'e, France}
\author{E. Quealy} \affiliation{Department of Physics, University of California, Berkeley} \affiliation{Physics Department, Napa Valley College}
\author{G. Rebeiz} \affiliation{Department of Electrical and Computer Engineering, University of California, San Diego}
\author{C.L. Reichardt} \affiliation{Department of Physics, University of California, Berkeley}
\author{P.L. Richards} \affiliation{Department of Physics, University of California, Berkeley}
\author{C. Ross} \affiliation{Department of Physics and Atmospheric Science, Dalhousie University, Halifax, NS, B3H 4R2, Canada}
\author{K. Rotermund} \affiliation{Department of Physics and Atmospheric Science, Dalhousie University, Halifax, NS, B3H 4R2, Canada}
\author{I. Schanning} \affiliation{Department of Physics, University of California, San Diego}
\author{D.E. Schenck} \affiliation{Center for Astrophysics and Space Astronomy, University of Colorado, Boulder} \affiliation{Department of Astrophysical and Planetary Sciences, University of Colorado, Boulder}
\author{B.D. Sherwin {\footnote{Corresponding author: \href{mailto: sherwin@berkeley.edu}{sherwin@berkeley.edu}}}} \affiliation{Department of Physics, University of California, Berkeley} \affiliation{Miller Institute for Basic Research in Science, University of California, Berkeley}
\author{A. Shimizu} \affiliation{The Graduate University for Advanced Studies}
\author{C. Shimmin} \affiliation{Department of Physics, University of California, Berkeley}
\author{M. Shimon} \affiliation{School of Physics and Astronomy, Tel Aviv University, Tel Aviv, Israel} \affiliation{Department of Physics, University of California, San Diego}
\author{P. Siritanasak} \affiliation{Department of Physics, University of California, San Diego}
\author{G. Smecher} \affiliation{Three-Speed Logic, Inc.}
\author{H. Spieler} \affiliation{Physics Division, Lawrence Berkeley National Laboratory}
\author{N. Stebor} \affiliation{Department of Physics, University of California, San Diego}
\author{B. Steinbach} \affiliation{Department of Physics, University of California, Berkeley}
\author{R. Stompor} \affiliation{AstroParticule et Cosmologie, Univ Paris Diderot, CNRS/IN2P3, CEA/Irfu, Obs de Paris, Sorbonne Paris Cit\'e, France}
\author{A. Suzuki} \affiliation{Department of Physics, University of California, Berkeley}
\author{S. Takakura} \affiliation{Osaka University} \affiliation{High Energy Accelerator Research Organization (KEK)}
\author{A. Tikhomirov} \affiliation{Department of Physics and Atmospheric Science, Dalhousie University, Halifax, NS, B3H 4R2, Canada}
\author{T. Tomaru} \affiliation{High Energy Accelerator Research Organization (KEK)}
\author{B. Wilson} \affiliation{Department of Physics, University of California, San Diego}
\author{A. Yadav} \affiliation{Department of Physics, University of California, San Diego}
\author{O. Zahn} \affiliation{Physics Division, Lawrence Berkeley National Laboratory}


\begin{abstract}
We reconstruct the gravitational lensing convergence signal from Cosmic Microwave Background (CMB) polarization data taken by the \Pb\ experiment and cross-correlate it with Cosmic Infrared Background (CIB) maps from the \herschel\  satellite. 
From the cross-spectra, we obtain evidence for gravitational lensing of the CMB polarization at a statistical significance of 4.0$\sigma$ and evidence for the presence of a lensing $B$-mode signal at a significance of 2.3$\sigma$. 
We demonstrate that our results are not biased by instrumental and astrophysical systematic errors by performing null-tests, checks with simulated and real data, and analytical calculations. 
This measurement of polarization lensing, made via the robust cross-correlation channel, not only reinforces \Pb\ auto-correlation measurements, but also represents one of the early steps towards establishing CMB polarization lensing as a powerful new probe of cosmology and astrophysics.\end{abstract}
\ifwordcount
\else
\maketitle
\fi

\textbf{Introduction:}
Precise measurements of the Cosmic Microwave Background (CMB) anisotropies with experiments such as WMAP, ACT, SPT, and Planck \citep{wmap1para,Das:2010ga,Keisler:2011aw,Planck:2006aa} have provided great insight into the evolution and composition of the Universe, yet a wealth of cosmological information remains undiscovered within the CMB. 
In particular, upcoming measurements of the gravitational lensing of the CMB -- deflections of CMB photons by the gravitational influence of the large-scale mass distribution -- 
are expected to make significant contributions to cosmology, probing the properties of dark energy \citep{sherwin11,vanEngelen:2012va}, the masses of neutrinos \citep{plancklensing}, and, through cross-correlations, the relation between dark matter and luminous tracers \citep{Smith:2007rg,Hirata:2008cb,Bleem:2012gm, feng12, actcross}.

The CMB lensing signal, which directly probes the mass distribution, can be estimated from lensing-induced correlations between CMB modes. Measurements of CMB lensing using temperature fluctuations have progressed from first detections \citep{Smith:2007rg,Hirata:2008cb,Das:2011ak,Feng:2011jx, vanEngelen:2012va,Bleem:2012gm, feng12, actcross} to precise measurements which constrain cosmological models \citep{plancklensing, planckCIB}. 
Further large improvements in precision and scientific return are expected from the measurement of lensing in the polarization of the CMB. 
Polarization lensing measurements can map the mass distribution in unprecedented detail, because, unlike temperature lensing, polarization measurements are not limited by cosmic variance.

Recent studies have measured the cross-correlation of the flux of the Cosmic Infrared Background (CIB) and CMB temperature lensing \citep{planckCIB,sptcib}. High correlation between the CMB lensing and CIB fields was found, with a maximal correlation coefficient of $\sim 80\%$ observed at a CIB wavelength of approximately $\sim 500 \mu$m.

Here we measure CMB polarization lensing and lensing $B$-mode polarization via a cross-correlation of \Pb{} CMB polarization lensing maps with maps of the CIB from \herschel, and verify that the signal agrees with theoretical expectation. Our work also demonstrates the novel technique of polarization lensing reconstruction in practice and supports the direct detection of the polarized lensing power spectrum by \Pb{} \citep{chang13} with an independent measurement.
Recently, a similar cross-correlation result was published by the SPT
collaboration \citep{hanson2013}. Our work differs in some aspects (for
example, our CMB maps have lower noise on a smaller area), but we find consistent results.
This agreement builds confidence in both results, given
the potentially different systematic errors (e.g., due to differences
in map depth, scan strategy, observing location or
experiment design).

\textbf{CMB and CIB Data:}
The \Pb\ experiment consists of a bolometric receiver operating on the 3.5-meter Huan Tran Telescope at the James Ax Observatory in Northern Chile \citep{Kermish_SPIE2012}.  The receiver has 1,274 polarization-sensitive transition-edge sensor bolometers observing a spectral band centered at 148 GHz \citep{Arnold_SPIE2012}.

This analysis uses data from two of the three fields observed between May 2012 and June 2013 (the two that overlap with \herschel\ data). 
The fields are roughly 10 sq. degrees in size and centered at (RA,Dec) = (23h02m, $-32.8^{\circ}$), (11h53m, $-0.5^{\circ}$), with approximate noise levels in polarization of 6$\mu$K-arcmin and 8$\mu$K-arcmin.
The fields will be referred to as ``RA23" and ``RA12," respectively.
The observations and map-making are described in the lensing power spectrum companion paper \citep{chang13}.

We construct an apodization window from a smoothed inverse variance weight of the \Pb{} map. 
Map pixels within 3\arcmin\ of ATCA catalog sources \cite{AT20G} are also masked. This catalog is measured at a frequency of 20 GHz. There are 12 sources masked from the two \Pb\ fields used.
We multiply the $Q$ and $U$ maps by this apodization window before transforming them into $E$ and $B$ maps using the pure-B transform \cite{Smith:2006}.

We also use overlapping 500\,$\mu$m data from the H-ATLAS survey \citep{eales2010} with the \herschel/SPIRE instrument \citep{pilbratt:2010}.
We  follow the approach of \citep{pascale2011} to create maps from the data, and use the maximum-likelihood map-maker HIPE \citep{hipe}, a calibration from \cite{swinyard2010} and standard pointing information. 
We flag glitches using the standard pipeline and replace them with constrained white noise realizations. Steps in DC-offset are compensated for by shifting the affected timelines
by an appropriate data-estimated DC-offset. The maps have rms instrument noise levels (per beam) of 7 mJy,
and the instrument has a Gaussian effective beam with FWHM 36.6\arcsec.

\textbf{Polarization Lensing Reconstruction and Cross-correlation Pipeline:}
CMB polarization is commonly described by two fields: an even parity $E$-mode polarization and an odd parity $B$-mode polarization field \citep{Kamionkowski:1996zd,zalsel}. Gravitational lensing by large-scale structure results in a remapping of the CMB photons described by the lensing deflection field $\mathbf{d}$, which points from the direction of photon reception to the direction of origin. Lensing converts $E$-modes into $B$-modes, inducing a correlation between the lensing $B$-modes and $E$-modes; similar correlations are also introduced between formerly independent pairs of $E$-modes. 

The optimal polarized quadratic estimators derive lensing by measuring these lensing-induced mode correlations \citep{Hu:2001tn, Hu:2001fa, HO}. The so-called $EB$ and $EE$ estimators are given by:
\beq
\hat{\mathbf{\kappa}}_{EB}(\mathbf L) = \int \frac{\mathrm{d}^2 \mathbf{l}}{(2 \pi)^2} {g}^{EB}(\mathbf{L},\mathbf{l}) E(\mathbf l) B(\mathbf{L}-\mathbf{l}),
\eeq
\beq
\hat{\mathbf{\kappa}}_{EE}(\mathbf L) = \int \frac{\mathrm{d}^2 \mathbf{l}}{(2 \pi)^2} {g}^{EE}(\mathbf{L},\mathbf{l}) E(\mathbf l) E(\mathbf{L}-\mathbf{l}),
\eeq
where $g$ is a function chosen as in \citep{HO} to normalize and optimize the estimator, $\mathbf{L}$ and $\mathbf{l}$ are Fourier space vectors conjugate to position on the sky, and $\kappa = - \nabla \cdot \mathbf{d}/2$ is the lensing convergence. Using these estimators, we calculate a noisy map of the lensing convergence field $\kappa$ which can be correlated with the \herschel\ CIB maps. In the estimators, we use only scales $500 \leq \ell \leq 2700$ in the polarization maps. This range of scales is chosen to ensure that the noise is effectively white (non-white noise increases at very low $\ell$) and that beam systematics \citep{Miller:2008zi} and astrophysical foregrounds \citep{vanEngelen:2012va} are subdominant (both increase at high $\ell$), while maintaining much of the possible signal-to-noise.

To test the pipeline, we generate a set of 400 Monte Carlo simulations which have similar properties to the data. To construct these simulations, we lens Gaussian simulations of CMB $Q$ and $U$ polarization using the method described in \citep{louis13}. We then add noise with the same level and spatial inhomogeneity as found in the data, with a constant power spectrum. Q/U noise correlations in the POLARBEAR data are only of order $1\%$; as the Q/U noise correlations are so small, and as they cannot bias a cross-power with large scale structure, we neglect them in our simulations. We verified that the deviation of the map noise power from white noise was minimal over the range of scales used in our analysis. 

These simulations are used to validate our pipeline as follows. We cross-correlate the reconstructed lensing convergence maps with the input lensing convergence maps in the simulation, which act as a proxy for the correlated part of the \herschel\ maps. By testing whether the resulting cross-power agrees with the noiseless lensing convergence power spectrum, we verify that our pipeline is unbiased. We repeat this pipeline validation with 100 CMB polarization signal simulations that have passed through the entire scanning and mapmaking pipeline, including the same source masking and window functions as used in the analysis of the data; with these simulations we verify that our estimators are correctly normalized to better than $\sim 15\%$ levels for both $EB$ and $EE$ estimators and are thus not significantly biased by scanning or mapmaking (this accuracy suffices for an approximate check of scan/mapmaking-effects to complement our main simulations, for which we have demonstrated much higher accuracy).

We use the lensing convergence maps, reconstructed as described previously, to measure the polarization lensing-CIB cross-power.

\textbf{Predicted Cross-power:}
As shown e.g. in \cite{Peiris:2000}, the cross-power is given by 
\begin{equation}
C_\ell^{\kappa I} = \int \frac{dz H(z)}{ \eta^2(z)} W^\kappa(z) W^I(z) P(k = \ell/\eta(z),z) 
\end{equation}
where $P(k,z)$ is the matter power spectrum, $W^I(z)$ is proportional to the redshift origin of the CIB signal $dI / dz$ and $W^\kappa$ is the CMB lensing kernel defined as in \citep{actcross}. 

We base our fiducial signal calculation of the lensing-CIB cross-power on the best-fit $W^I(z)$ at $500\mu$m from \cite{sptcib}, which in turn relies on the model of \citep{viero2013}. The resulting signal theory curve is used in Figs. 1 and 2.

\textbf{Measured Cross-power:}
We measure the cross-powers of polarization lensing and the \herschel\ maps of the infrared background on two \Pb\ maps (RA12 and RA23), with lensing derived from both the $EB$ and $EE$ estimators. All four cross-power spectra (two estimators on two maps) are shown in the lowest panel of Fig.~1. We co-add the two cross-spectra involving the $EB$ estimator to calculate a cross-power corresponding to a measurement of $B$-mode polarization, shown in the middle panel of Fig.~1; we obtain evidence for $B$-mode polarization from lensing at a significance of 2.3$\sigma$. The significance of a detection is calculated using the expression $\sqrt{\sum_i (\chi^2_{i,\mathrm{null}}-\chi^2_{i,\mathrm{theory}})}$ where the sum is over all relevant cross-powers and $\chi^2$ is calculated using the full covariance matrix. We similarly construct a co-added combination of all four polarized lensing-cross powers, shown in the top panel of Fig.~1; this corresponds to a detection of polarized lensing at 4.0$\sigma$ significance. 

\begin{figure}[p]
\label{fig.crosspower}
    \includegraphics[width=3.75in]{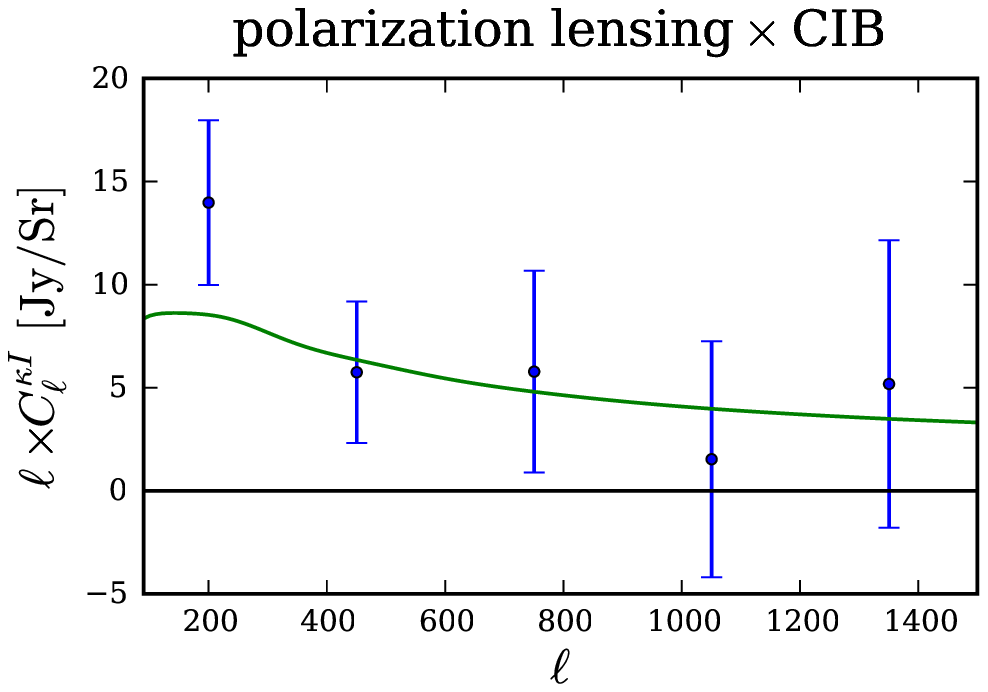}
    \includegraphics[width=3.75in]{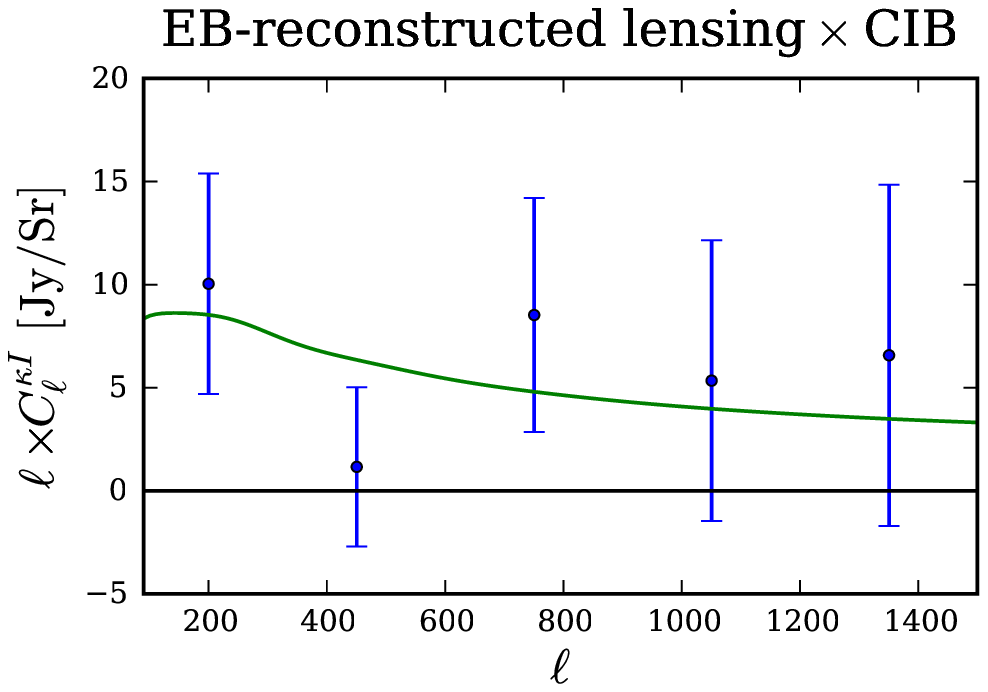}
    \includegraphics[width=3.75in]{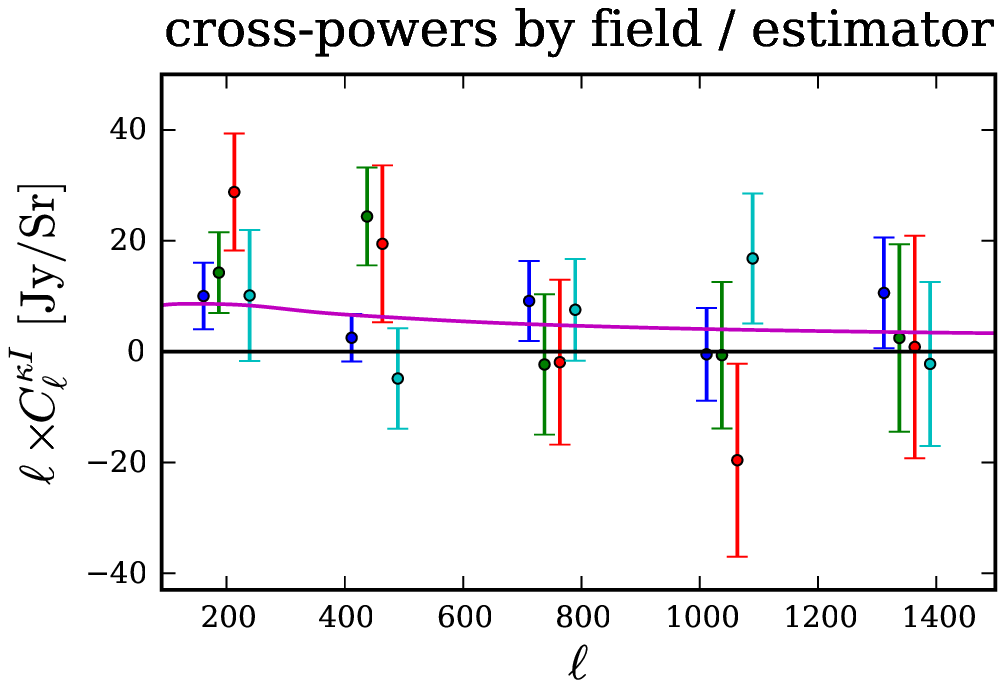}
    \caption{Cross-power spectra of CMB polarization lensing and the 500$\mu$m \herschel\ CIB flux. {\bf Top panel}: the minimum variance combination of all polarization lensing measurements cross-correlated with the \herschel\ maps; this result corresponds to 4.0$\sigma$ evidence for gravitational lensing of CMB polarization. {\bf Middle panel}: the cross power of $EB$-reconstructed lensing with the \herschel\ maps; constructed from the $EB$ estimator applied to both \Pb\ maps, this result corresponds to 2.3$\sigma$ evidence for lensing $B$-modes.  {\bf Bottom panel}: all four combinations of the two lensing estimators ($EE, EB$) applied to two different \Pb\ maps (RA23, RA12) and cross-correlated with \herschel\ - $EB$/RA23 (dark blue), $EE$/RA23 (green), $EE$/RA12 (red), $EB$/RA12 (cyan), listing from left to right for each bandpower. The fiducial theory curve for the lensing -- CIB cross-spectrum \cite{sptcib} is also shown (solid line).}
\end{figure}

The errors and the 5 $\times$ 5 covariance matrix for our cross-power measurement are obtained using the 400 simulations described earlier. We perform simple convergence tests by varying the number of simulations used and find stable results. 
We also note that the errors we simulate agree with the results of analytical calculations based on the observed power spectra and that using only the diagonal elements of the covariance matrix gives similar detection significances (to within 0.2$\sigma$).  
As our null hypothesis is the absence of gravitational lensing of CMB polarization, we include no lensing in the simulations used to derive the detection significance.

\textbf{Systematic Error Estimates and Null Tests:}
Here we discuss the effects of potential sources of systematic error on the polarization lensing -- CIB cross-correlation. 
We first focus on astrophysical foregrounds before turning to instrumental systematics. 
To check the point source contamination level, we compare the lensing-CIB cross-powers with and without the ATCA sources masked in the CMB; we find the differences are less than 0.2$\sigma$, which indicates that the contribution of polarized point sources is negligible. As an additional test, we simulate polarized radio point sources in both CMB and CIB maps and propagate these maps through our lensing estimation and cross-correlation pipeline. We very conservatively estimate 10\% polarization fraction, counts as in \citep{cmbpol}, and neglect any source masking in \Pb. We find negligible contamination to the cross-power, at levels well below a percent of the signal. 

We next consider contamination due to polarized emission from dusty CIB sources, which could potentially propagate through the lensing estimator and bias our cross-correlation measurement. We first construct 400 simple simulations of this effect, approximating the CIB maps as Gaussian random fields with the  \herschel\ 500$\mu$m power spectrum (a good approximation because source-confusion dominates). Polarized CIB $Q$ and $U$ simulations are obtained by rescaling these maps, assuming a frequency dependence as in \cite{planckCIB} and very conservatively assuming each pixel has a randomly-oriented averaged polarization fraction of $3\%$. With these simulations, we find a bias to the cross-power consistent with zero (less than $2 \times 10^{-4}$ of the expected cross-power signal), as expected from a Gaussian three-point function, and a negligible change in the error bars (less than $7\%$). Though Gaussian simulations are a good approximation to the \herschel\ maps, we also test for contamination from the brightest, Poisson-distributed polarized dusty sources. In our test, we compare the signals with and without the brightest regions in the \herschel\ maps masked (where the flux in a $2 \times 2$ arcmin$^2$ pixel is greater than that of a 50mJy source averaged over the pixel). We find changes to the signal at the level of only 0.2 $\sigma$, indicating bright infrared source contamination is negligible.
\comment{We note that no other systematics can mimic the galaxy $EB$-lensing cross-correlation signal, as the signal will still be zero on average if the $B$-mode map contains no lensing information. } 

We next discuss instrumental systematic errors. 
First, we consider a general systematic that linearly couples $T$- and $E$-modes into $B$-modes, as leakage most affects the small $B$-mode signal. To estimate the effects of such instrumental systematic errors, we simply insert a general expression for the systematic-contaminated $B$-mode
\beqn
\tilde{B}(\mathbf{l}) &=& B (\mathbf{l}) + \int \frac{\mathrm d^2 \mathbf{l}'}{(2\pi)^2} s_{EB}(\mathbf{l}-\mathbf{l}')E(\mathbf{l}') \nonumber \\ &+& \int \frac{\mathrm d^2 \mathbf{l'}}{(2\pi)^2} s_{TB}(\mathbf{l}-\mathbf{l}')T(\mathbf{l}'),
\eeqn
into our expression for the cross-correlation using the $EB$-reconstruction.
Here the functions $s$ describe the systematic-induced couplings and the fields $E, B, T$ are the true (lensed) fields on the sky. 
Analytically calculating the effects of such leakage on the cross-correlation, we find that the bias it induces is zero (to first order in $s$). This is due to the fact that, in cross-correlation, the $EB$-estimator is insensitive to leakage of even parity.
To test this analytic calculation in simulations, we repeat the cross-correlation pipeline verification described earlier, except now introducing leakage terms. We add 1\% of the temperature maps to the $Q$ and $U$ maps, and add 10\% of $Q$ to $U$ and vice versa. The introduced leakage does not bias the cross-correlation to percent-level accuracy although the errors increase marginally. We perform a similar simulated test of the effect of leakage on the cross-power calculated with the $EE$ estimator. Using the same simulations of systematic leakage as for the $EB$ estimator, we again find a negligible ($2\%$) change in the recovered cross-power.

We estimate the effects of beam uncertainty by generating simulations with the beam values everywhere increased or decreased by an amount equal to the $1\sigma$ error. Despite using a coherent offset across all maps and scales, we find only small effects, always significantly less than $20\%$ of the signal.  Differential beam ellipticity results in leakage of temperature to polarization \citep{Shimon08}; however, as investigated with both analytical arguments and simulations, such leakage does not bias our results. Common-mode beam ellipticity is expected to be highly subdominant as the scan-pattern on our maps is nearly isotropic. As described in \citep{chang13}, we also constrain the gain error to be less than $10\%$ of the signal. However, we note that such beam and gain errors cannot mimic a detection in cross-correlation if there is in fact no polarization lensing signal. Finally, we note that our results are insensitive to $\approx 1$ degree polarization angle rotation, with the resulting shifts typically below $0.2\sigma$. In comparison, the statistical error on our angle measurement is $\approx 0.2$ degrees. This again confirms the insensitivity of our cross-correlation to possible sources of instrumental systematic error.

We further verify our cross-correlation measurement with a number of null tests. First, we cross-correlate the \herschel\ maps with non-overlapping \Pb\ maps -- calculating for instance \Pb\ RA12 $\times$ \herschel\ RA23 and \Pb\ RA23 $\times$ \herschel\ RA12 (as in \citep{Das:2011ak}). 
Failed null tests would indicate that our simulated error bars have been underestimated. 
The results are seen in Fig. 2 -- the null tests are passed for both patches of sky, with $\chi^2$ probabilities-to-exceed (PTEs) of 42\%/49\% ($EB$ estimator / $EE$ estimator) for \Pb\ RA12 $\times$ \herschel\ RA23 and 24\%/56\% ($EB$ estimator / $EE$ estimator) for \Pb\ RA23 $\times$ \herschel\ RA12 respectively.

\begin{figure}[h]
\label{fig.filter}
  \includegraphics[width=3.75in]{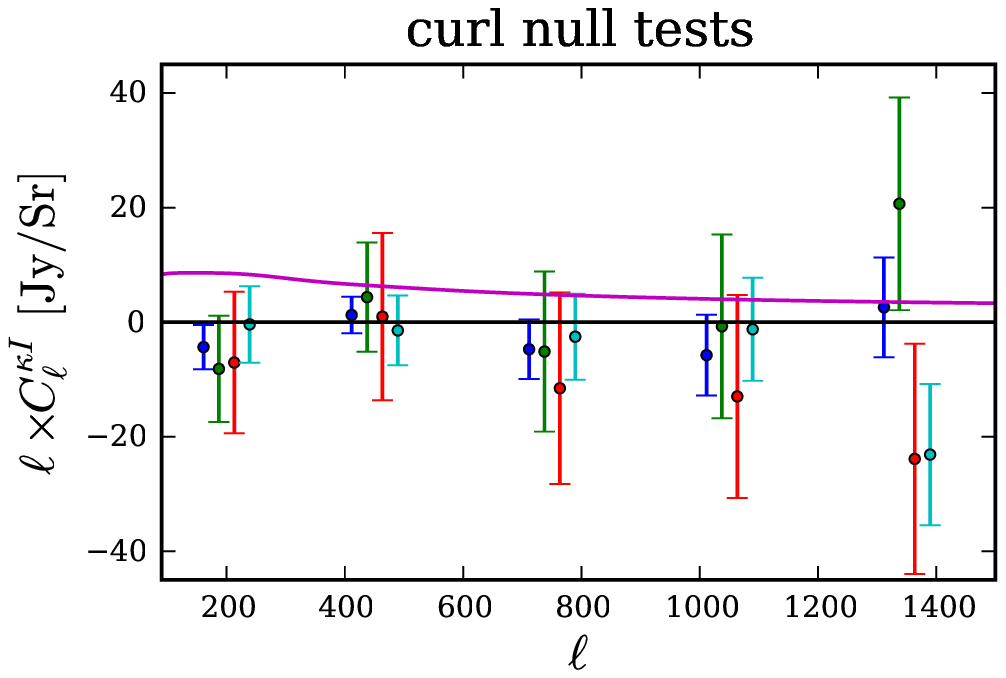} 

    \includegraphics[width=3.75in]{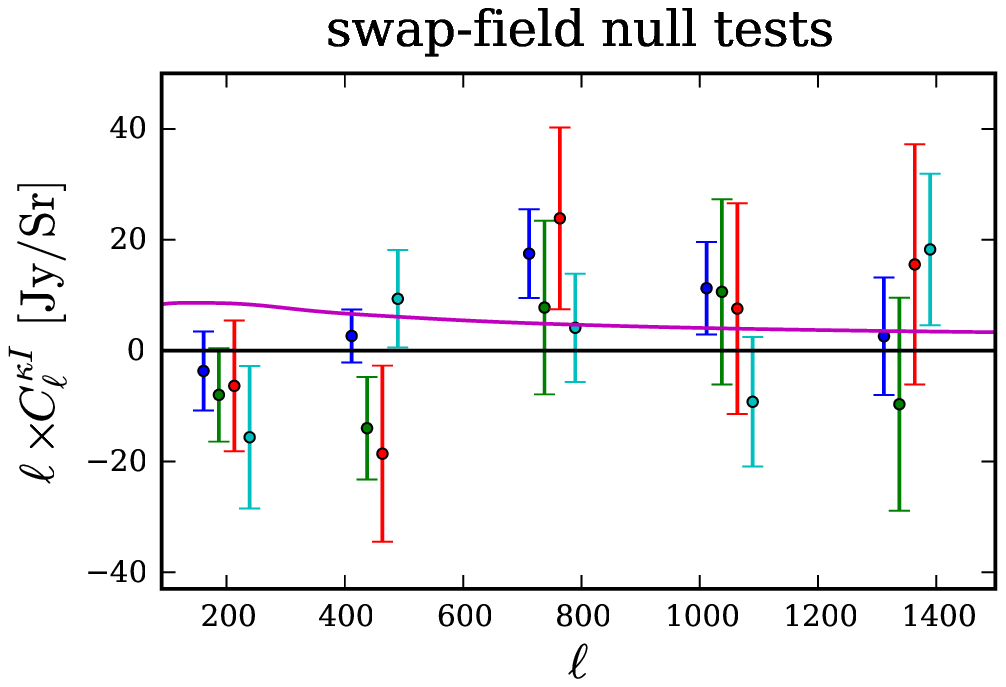}      
    \caption{
    {\bf Upper panel}: Curl null tests for all four combinations of estimator / map -- $EB$/RA23 (dark blue), $EE$/RA23 (green), $EE$/RA12 (red), $EB$/RA12 (cyan). In this figure, curl null test values and errors have been scaled down by a factor of two for the $EE$ estimator, for ease of plotting (PTEs are unaffected).
    {\bf Lower panel}: Swap-field null tests. Note that the four sets of points are not entirely statistically independent. The null tests are consistent with zero and thus provide no evidence for any systematic errors.}
\end{figure}

In a second null test, we calculate the curl component of the lensing deflection field (as in \citep{vanEngelen:2012va}) and cross-correlate it with the \herschel\ map on the same patch of sky. 
The curl null test can probe sources of systematic leakage of temperature and $E$-modes into $B$-mode polarization such as pixel rotation or differential effects in gain, beamwidth, and ellipticity, as the resulting leakage has different parity properties which can be probed by the curl estimator.
The results of the curl null test are shown for both patches in Fig 2. The PTEs for this null test are 74\%/81\% for the RA23 map (for the $EB$ and $EE$ estimators, respectively) and 55\%/72\% for the RA12 map (again for the $EB$ and $EE$ estimators). The curl null test results are consistent with zero.

Finally, we note that a large number of systematic checks and null tests are performed in \citep{chang13}, with the same \Pb\ CMB data and maps, for measurements of both the lensing and polarization power spectra. 
Though both polarization and lensing power spectrum measurements are much more sensitive to systematic errors than a cross-correlation measurement, no evidence for systematic contamination is found in either case. This gives further confidence in the robustness of our results.

To prevent observer bias, we used a blind analysis for the $B$-mode portion of our results: $B$-mode cross-spectra were kept blinded until all systematic tests had been performed.


\textbf{Conclusions:}
We report evidence for polarized lensing at 4.0$\sigma$ and evidence for lensing $B$-modes at 2.3$\sigma$ significance, from a measurement of the CMB lensing -- CIB cross correlation. 
This measurement is robust against both astrophysical and instrumental systematic errors, and is hence a particularly reliable measurement of polarized lensing. 
Our results thus reinforce the detection of the polarization lensing power spectrum reported in \citep{chang13}. 
This work and other measurements of polarization lensing lie at the beginning of an exciting new field, which will survey the high-redshift mass distribution in detail, provide powerful constraints on the properties of dark energy, neutrinos and inflation, and give insight into the relation between dark and luminous matter.

\ifwordcount
\else
\acknowledgements
\textbf{Acknowledgments:}
We thank Frank Wuerthwein, Igor Sfiligoi, Terrence Martin, and Robert Konecny for their insight and support, and thank Nolberto Oyarce and Jos\'{e} Cortes for their invaluable
contributions. Calculations were performed at the Department of Energy Open Science Grid \citep{Pordes2008} at the University of California, San Diego, accessed via the GlideinWMS \citep{Sfiligoi2009}, at Central Computing System, owned and operated by the Computing Research Center at KEK, and at NERSC which is supported by the DOE under Contract No. DE-AC02-05CH11231. 
The \Pb{} project is funded by the NSF under grant AST-0618398 and AST-1212230.
The KEK authors were supported by MEXT KAKENHI Grant Number 21111002, and acknowledge support from KEK Cryogenics Science Center.
The McGill authors acknowledge funding from the Natural Sciences and Engineering Research Council and Canadian Institute for Advanced Research. 
NM, BDS, and KA acknowledge support from the NASA Postdoctoral Program, a Miller Fellowship, and the Simons Foundation, respectively. MS gratefully acknowledges support from Joan and Irwin Jacobs. 
The James Ax Observatory operates in the Parque Astron\'{o}mico
Atacama in Northern Chile under the auspices of the Comisi\'{o}n Nacional de Investigaci\'{o}n Cient\'{i}fica y Tecnol\'{o}gica de Chile (CONICYT). Finally, we acknowledge the tremendous contributions by Huan Tran to the \Pb\ project.
\comment{
\begin{appendix}

\section{Instrumental Systematics in the Lensing-tracer Cross-correlation}
In this Appendix, we focus on systematics for the cross-correlation of the $EB$ estimator with the CIB.
The cross-correlation of a tracer $I$ with $EB$-reconstructed lensing is given by the following equation
\beq
\label{ap2}
C^{\kappa I}_L =   \int  \frac{d^2\Bell}{(2\pi)^2}  g^{EB}(\Bell,\bL) ~  \langle  I^*(\bL)E(\Bell) \tilde B(\bL -\Bell)  \rangle
\eeq
where $g$ is an optimal weight function multiplied by a normalization $N(L)$:

\beq
 g^{EB}(\Bell,\bL) = N^\kappa(L) \frac{\Bell \cdot \bL ~C^{EE}_\ell \sin(2 \phi _{\Bell,\bL-\Bell})}{C^{BB}_{|\bL-\Bell |}+N^{BB}_{|\bL-\Bell |}} \frac{C^{EE}_\ell}{C^{EE}_\ell+N^{EE}_\ell} 
\eeq

We will approximate $\tilde E = E$ for our purposes (where tildes indicated quantities including systematic $T/E/B$ leakage), as we assume the systematic contribution to $E$-modes is known to be negligible from accurate measurements of $C_\ell^{EE}$, which are affected by any such systematic. In addition, lensing cross-correlations using the $EB$ estimator are zero without the presence of lensing information in the $B$-modes, regardless of whether or not $E$ is contaminated by systematics. The most worrying systematics for detecting $B$-modes through $EB$-reconstructed lensing cross-correlation are ones which propagate spurious lensing information into the $B$-modes; we can thus focus on systematics $s_{EB}$ and $s_{EE}$ which couple $E$-mode and temperature power into $B$-modes, and contain the (small) cosmological lensing signal, thus potentially inducing a spurious correlation with high-redshift tracers.

The systematic-contaminated $B$-mode signal in Eq.~5 has three terms: the true $B$-mode, the part from systematic coupling of $E$ to $B$, and the part from systematic coupling of $T$ to $B$. The estimated cross power thus also splits into these terms (by linearity in the $B$-modes), so we can write
\beq
C_L = C^{\mathrm{true}}_L + C^{s_{EB}}_L + C^{s_{TB}}_L
\eeq
where $s$ describes the systematic mode coupling as in Eq.~5.

We can now insert each of the terms of Eq.~5 into Eq.~B1 and use the expressions from Hu and Okamoto (2002) for $\langle E(\Bell) X(\bL - \Bell) \rangle_{\mathrm{CMB}} = K_{EX}(\Bell, \bL) \kappa(\bL) \times 2 / L^2$, where $X = T, E, B$. We thus obtain the expression for the contamination of the lensing by $E$ to $B$ leakage systematics:

\beqn
C_L^{s_{EB}} &=& s_{EB}(0) C_L^{\kappa I} \int  \frac{d^2\Bell}{(2\pi)^2} g^{EB}(\Bell, \bL)\\ \nonumber &\times& \cos(2 \phi _{\Bell,\bL-\Bell})[(\bL-\Bell) \cdot \bL ~C^{EE}_{|\bL-\Bell |} +\Bell \cdot \bL ~C^{EE}_\ell ]\eeqn

Similarly, we obtain for $T$ to $B$ leakage systematics:

\beqn
C_L^{s_{TB}} &=& s_{TB}(0) C_L^{\kappa I} \int  \frac{d^2\Bell}{(2\pi)^2}  g^{EB}(\Bell, \bL)\\ \nonumber &\times& [(\bL-\Bell) \cdot \bL  ~C^{TE}_{|\bL-\Bell |} \cos({2 \phi _{\Bell,\bL-\Bell}}) + \Bell \cdot \bL  ~C^{TE}_{\ell}]\eeqn

To assess the magnitude of these three $C_L$ terms, we consider how the functions $g^{EB}, K_{EB}, K_{EE}, K_{TE}$ behave under a tranformation ``reflecting'' $\Bell$ about $\bL$. First, we note that $g^{EB}$ is odd under the reflection of $\Bell$ about the direction $\bL$. While $K_{EB}$ is odd under this reflection, $K_{EE}$ and $K_{TE}$ are even; the products $g^{EB} K_{EE}$ and $g^{EB} K_{TE}$ are thus odd functions about the direction of $\bL$ so that $C_L^{s_{EB}} =C_L^{s_{TB}} = 0$. This result can be simply interpreted by noting that the optimal $EB$-estimator is designed to be maximally sensitive to the odd parity $B$-modes and hence is insensitive to leakage from temperature and $E$-modes which have even parity. We verify the analytical arguments for the integrals being zero by calculating them numerically, and find results consistent with zero to within numerical precision.

One could imagine caveats to the arguments above, if mode coupling or anisotropic filtering prevented the complete zeroing of the integrals giving the systematic contributions, or if the systematic contribution to the $E$-mode were unexpectedly large. However, even in this case we can argue that systematics should be very small in cross-correlations, as long as they do not completely dominate the polarization power spectra. For this, let us consider the contribution of systematics to the power spectra:

\beqn
\langle \tilde B^*(\Bell) \tilde B(\Bell) \rangle &=& C^{BB}_\ell + \int \frac{\mathrm d^2 \Bell'}{(2\pi)^2} [ | s_{EB}(\Bell-\Bell')|^2 C^{EE}_{\ell'} \\ \nonumber &+&  | s_{TB}(\Bell-\Bell')|^2 C^{TT}_{\ell'} + 2 \mathrm{Re}(s^*_{EB} s_{TB})(\Bell-\Bell')C^{TE}_{\ell'} ] 
\eeqn

\beqn
\langle \tilde E^*(\Bell) \tilde E(\Bell) \rangle &=& C^{EE}_\ell + \int \frac{\mathrm d^2 \Bell'}{(2\pi)^2} \left[ | s_{TE}(\Bell-\Bell')|^2 C^{TT}_{\ell'}  \right]  \nonumber \\ &+&  2 \mathrm{Re}(s^*_{TE}(0))C^{TE}_{\ell}
\eeqn

If the systematic contamination to the power spectra is to be subdominant, we require that the coefficients $s_{EB} << \sqrt{(C^{BB} / C^{EE} ) }$  and $s_{TB} << \sqrt{(C^{BB} / C^{TT} ) }$ . This is more than sufficient for the lensing terms also to be subdominant, as by a rough estimate they require only the weaker conditions $s_{EB} << 1$ and $s_{TB} << {(C^{EE} / C^{TE} ) }$ even if the integrals are not zero but are instead order unity. We have verified through tests with the polarized power spectra that the systematic contamination is not a dominant contribution. Similar arguments for why systematics should be smaller in cross-correlation than in auto-correlation also apply to the $EE$ lensing estimator.

We conclude that if we observe a non zero $EB$-reconstructed-lensing -- galaxy cross-correlation signal, it is unlikely to be caused by instrumental systematics of any type.

\end{appendix}
}

\bibliography{pb}
\fi

\end{document}